\documentclass[11pt,letter]{article}
\usepackage{deauthor}
\usepackage{times}
\usepackage{graphicx}
\usepackage{amsfonts}   
\usepackage{amsmath}
\usepackage{amssymb}
\usepackage{booktabs}
\usepackage{multirow}
\usepackage[font=small,labelfont=bf]{caption}

\usepackage{deauthor}

\usepackage[utf8]{inputenc} 
\usepackage[T1]{fontenc}    
\usepackage{hyperref}       
\usepackage{url}            
\usepackage{booktabs}       
\usepackage{amsfonts}       
\usepackage{nicefrac}       
\usepackage{microtype}      
\usepackage{lipsum}

\usepackage{graphicx}

\usepackage{graphicx}
\usepackage{algorithm}
\usepackage{caption}
\usepackage{algorithmicx}
\usepackage{algcompatible}

\begin{document}

\title{A Community-aware Network Growth Model for Synthetic Social Network Generation}

\def\sharedaffiliation{%
\end{tabular}
\begin{tabular}{c}}
\author{
Furkan Gursoy\\
\texttt{furkan.gursoy@boun.edu.tr}
\and
Bertan Badur \\
\texttt{bertan.badur@boun.edu.tr}
\vspace{10pt}
\sharedaffiliation
Dept. of Management Information Systems\\
  Boğaziçi University\\
  Istanbul, Turkey
}

\maketitle

\begin{abstract}
This study proposes a novel network growth model named \textit{ComAwareNetGrowth} which aims to mimic evolution of real-world social networks. The model works in discrete time. At each timestep, a new link (I) within-community or (II) anywhere in the network is created (a) between existing nodes or (b) between an existing node and a newcoming node, based on (i) random graph model, (ii) preferential attachment model, (iii) a triangle-closing model, or (iv) a quadrangle-closing model. Parameters control the probability of employing a particular mechanism in link creation. Experimental results on Karate and Caltech social networks shows that the model is able to mimic real-word social networks in terms of clustering coefficient, modularity, average path length, diameter, and power law exponent. Further experiments indicate that \textit{ComAwareNetGrowth} model is able to generate variety of synthetic networks with different statistics.
\end{abstract}


\section{Introduction}
Purpose of this study is to build a network growth model that is able to generate synthetic networks which resemble real-world social networks. Existing network growth models in the literature often aim to reproduce only a very small set of network statistics. Our work, on the other hand, attempts to reproduce a larger list of various statistics to evaluate the proposed model's accuracy.  

The contribution of this study includes using combination of different growth mechanisms under a community-aware framework, providing a growth model to understand evolution of many social networks, and providing a generative mechanism which can be used in generating synthetic networks.

The rest of this study is structured as follows. In Section 2, the relevant literature is briefly reviewed. In Section 3, the proposed model is formally described. In section 4, experimental results and findings are presented. The conclusion and final remarks are given in Section 5.

\section{Background}

Generally speaking, random graph models assume that the probabilities of links (edges) between nodes (vertices) are independent of each other and equal. There are closely related versions of random graph models. For instance, in original Erdos-Renyi model (Erdos \& Renyi, 1959), all graphs on a fixed node set with a fixed number of links are equally likely. On the other hand, in the model introduced by Gilbert (1959), each link has a fixed probability of being present or absent, independently of the other links.

Random graph models and many other popular models in the literature are originally static models which enable the study of structural features of networks. Another category of network models is growth models (also called as generative network models). Preferential attachment model is an example of a model in the latter category. (Simon, 1955), (Price, 1976), (Barabasi \& Albert, 1999) are the seminal papers on preferential attachment model. The main idea behind this model is as follows. At each time, a new node arrives and makes a certain number of links with existing nodes using the node degrees as probability of making a link with the node. This mechanism is able to generate a degree distribution which follows a power law which is observed in most real-world networks.

Triadic closure is a concept in social network theory, first suggested by German sociologist Simmel (1908). It follows the idea that two nodes are more likely to be connected if they have common neighbors. Generalizing this, we can consider $n$-th order neighbors instead of only first-order neighbors. In this way, for instance, a quadrangle-closure implies that two nodes are more likely to be connected if they have neighbors who are neighbors with each other. Once a quadrangle is closed, a four-cycle is created. Lazega and Pattison (1999) examine whether cycles larger than tri-cycles could be observed in an empirical setting to a greater extent than could be accounted for by parameters for configurations involving at most three nodes. Snijders et al. (2006) describe this as four-cycle partial conditional dependence and proposes it as a new configuration.

Kimura et al. (2004) present a model where links are created by a mixture of preferential attachment, uniform attachment, and community-based attachment model. Leskovec et al. (2008) propose a network evolution model considering the micromechanisms in social networks. Their model utilizes preferential attachment and triangle-closure models in creating links. Lim et al. (2016) provide a comprehensive review of different approaches in generating realistic synthetic graphs.

\section{Proposed Model}
The proposed model works in discrete time steps. It assumes that each node belongs to exactly one community. At each time step, a new link is to be created. Nodes are assumed to arrive in uniform time intervals. The time interval is found by dividing number of links ($m$) to number of nodes ($n$), thus a new node arrives at each  $m/n$-th step. 

When a new node arrives, it is assigned a community label with respective probabilities given as input. Then, the arriving node makes a link either within its community or in the whole network based on respective probabilities specified by the input parameter. The link can be made with a uniformly selected random node or with a node selected by using node degrees as the proxy probabilities (i.e., preferential attachment). If the link is to be made within community, node degrees for preferential attachment process is calculated only considering the links in the community.

At the time steps where no new node arrives, a new link is created between the existing nodes. One end of the link is selected randomly among all existing nodes. The other end is selected within community/in whole network, randomly or based on preferential attachment model; similar to the procedure explained previously for a newly arriving node. However, another mechanism is available for links between existing nodes that is not available for newly arriving nodes: triangle or quadrangle closing models. For a given node, a triangle is closed by making a link to its second-order neighbor. A quadrangle is closed by making a link to its third-order neighbor. More mechanisms can be created based on the proposed $n$-th order specification but considering that diameter and average path length of observed networks are relatively small, including such mechanisms with higher $n$ value would mean making a link specifically to a distant node rather than a node in the neighborhood.

The selection of mechanisms (e.g., within community/in whole network, random/preferential attachment/triangle-closure/quadrangle-closure) are controlled by input parameters. In some cases, the selected mechanism is not able to generate a new link for reasons such as lack of possible triangles or quadrangles, or because all possible links for the given node and mechanism already exist in the network. In those cases, at that time step, no action is performed and model continues with the next time step. Although the property of uniform arrival of nodes is distorted, the resulting network still has $n$ nodes and $m$ links. 

In addition, potential target nodes are selected before checking whether they already have a link with the source node. If a selected node is already connected, no action is performed at that time step too. This also results in violation of uniform arrival of nodes. However, we chose to keep it this way to prevent overdensification of relatively small communities where a randomly selected node is more likely to be an immediate neighbor.

Each mechanism employed in \textit{ComAwareNetGrowth} model corresponds to the simplified versions of possible link formations in real-world networks. The stronger the communities, the more probable for a node to make connection within its community. Random links are usually less frequent in most social networks but actually an existing mechanism. Preferential attachment reflects the 'rich get richer' phenomena. Triangle closures and quadrangle closures are justified by the observation that nodes are usually more likely to make links within their neighborhood rather than distant nodes.

Algorithm 1 presents our proposed model in general with omitting some details. The complete source code in R software language and some network datasets generated using the proposed methodology are available at \href{https://furkangursoy.github.io}{\textit{furkangursoy.github.io}}.

\begin{algorithm}[ht]
\caption{\textit{ComAwareNetGrowth}: The Proposed Community-aware Network Growth Model}
\label{alg:ouralgorithm}
\centering
\includegraphics[width=0.97\linewidth]{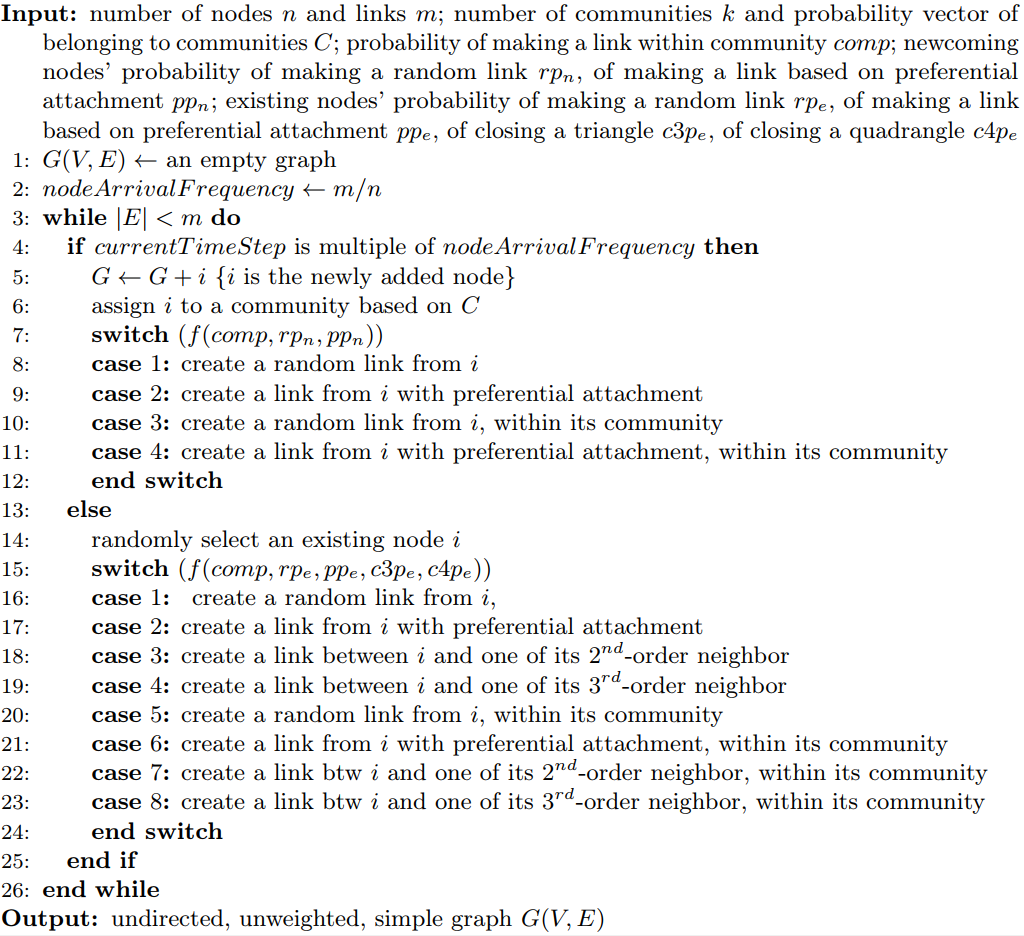}
\end{algorithm}

\section{Experimental Analysis}

In this section, several experiments are conducted and results are analyzed. In the first part, we attempt to generate two real-world undirected simple social networks: Zachary's Karate Club (Zachary, 1977) and Caltech Facebook Network (Traud et al., 2011). The first network, called as Karate in the rest of the study, is a well-known social network of a university karate club where links exist between members who interact outside the club. The latter network, called as Caltech in the rest of the study, consists of the complete set of users from the Facebook network of California Institute of Technology.

Table 1 displays various statistics of the two networks. The statistics are calculated using R's igraph package. In calculation of modularity; first, communities are detected by utilizing fast greedy community detection algorithm available in R. Estimated communities, then, are used in calculation of modularity. The sizes of the found communities are also utilized as input probabilities for belonging to communities in the network growth experiments.

\begin{table}[h!]
\centering
\caption{Network Statistics}
\small
\begin{tabular}{|l|r|r|r|r|r|r|r|} 
\cline{2-8}
\multicolumn{1}{l|}{} & n   & m     & \begin{tabular}[c]{@{}r@{}}Clustering\\ Coefficient \end{tabular} & \begin{tabular}[c]{@{}r@{}}Avg. Path\\ Length \end{tabular} & Modularity & Diameter & \begin{tabular}[c]{@{}r@{}}Power Law\\ Exponent \end{tabular}  \\ 
\hline
Karate                & 34  & 78    & 0.26                                                              & 2.41                                                        & 0.38       & 5.00     & 2.55                                                           \\ 
\hline
Caltech               & 769 & 16656 & 0.29                                                              & 2.34                                                        & 0.33       & 6.00     & 1.50                                                           \\
\hline
\end{tabular}
\end{table}

\newpage
\subsection{Experiments on Karate Network}

The following parameter values are used to generate a social network that mimics Karate network: $n = 34$, $m = 78$, $k = 3$, $C = \{0.24, 0.5, 0.26\}$, $comp = 0.8$, $rp_n = 0.22$, $pp_n = 0.78$, $rp_e =0.04$, $pp_e = 0.14$, $c3p_e = 0.41$, and $c4p_e = 0.41$. The experiments are repeated 10 times, and statistics for the generated and observed networks are given in Table 2.

On average, the proposed growth model with the given parameter settings is able to mimic the Karate network in general. However, modularity and and power law exponent\footnote{Most networks follow power-law in the tail. Accordingly, in fitting the power law, lower bound is set as 2 for Karate and 5 for all other experiments.} (Newman, 2005) statistics are not closely matched. Although most of the statistics can be generated with small standard deviations, modularity and clustering coefficient show relatively larger variations in these experiments. This might be due to the very small size of the network.

\begin{table}[h!]
\centering
\caption{Experimental Results for Karate Network}
\small

\begin{tabular}{|l|r|r|r|r|r|} 
\cline{2-6}
\multicolumn{1}{l|}{} & \begin{tabular}[c]{@{}r@{}}Clustering\\ Coefficient \end{tabular} & \begin{tabular}[c]{@{}r@{}}Avg. Path\\ Length \end{tabular} & Modularity      & Diameter       & \begin{tabular}[c]{@{}r@{}}Power Law\\ Exponent \end{tabular}  \\ 
\hline
Run\#1                 & 0.18                                                              & 2.32                                                        & 0.26            & 4.00           & 2.20                                                           \\ 
\hline
Run\#2                 & 0.24                                                              & 2.42                                                        & 0.26            & 5.00           & 2.13                                                           \\ 
\hline
Run\#3                 & 0.22                                                              & 2.35                                                        & 0.20            & 4.00           & 2.10                                                           \\ 
\hline
Run\#4                 & 0.25                                                              & 2.47                                                        & 0.27            & 5.00           & 2.31                                                           \\ 
\hline
Run\#5                 & 0.19                                                              & 2.50                                                        & 0.34            & 5.00           & 2.22                                                           \\ 
\hline
Run\#6                 & 0.30                                                              & 2.38                                                        & 0.27            & 5.00           & 2.20                                                           \\ 
\hline
Run\#7                 & 0.20                                                              & 2.31                                                        & 0.21            & 5.00           & 2.24                                                           \\ 
\hline
Run\#8                 & 0.27                                                              & 2.52                                                        & 0.22            & 6.00           & 2.02                                                           \\ 
\hline
Run\#9                 & 0.30                                                              & 2.55                                                        & 0.22            & 6.00           & 2.14                                                           \\ 
\hline
Run\#10                & 0.27                                                              & 2.56                                                        & 0.34            & 5.00           & 2.19                                                           \\ 
\hline
\textbf{Mean}         & \textbf{0.24}                                                     & \textbf{2.44}                                               & \textbf{0.26}   & \textbf{5.00}  & \textbf{2.18}                                                  \\ 
\hline
\textit{StdDev}       & \textit{0.04}                                                     & \textit{0.09}                                               & \textit{0.05}   & \textit{0.63}  & \textit{0.08}                                                  \\ 
\hline
\textbf{Observed}     & \textbf{0.26}                                                     & \textbf{2.41}                                               & \textbf{0.38}   & \textbf{5.00}  & \textbf{2.55}                                                  \\ 
\hline
\textbf{Diff.}        & \textbf{-0.02}                                                    & \textbf{0.03}                                               & \textbf{-0.12}  & \textbf{0.00}  & \textbf{-0.37}                                                 \\
\hline
\end{tabular}
\end{table}

Figure 1  visualizes the communities found in the original Karate network and a synthetic network generated by our model. In the synthetic network, communities are less separated than the original network. Especially, two communities (displayed in white and black) do not come as separate communities. Note that, this is just one of the generated networks. Other networks might more closely resemble the original network.

\begin{figure}[h!]
\centering
  \includegraphics[width=0.85\linewidth]{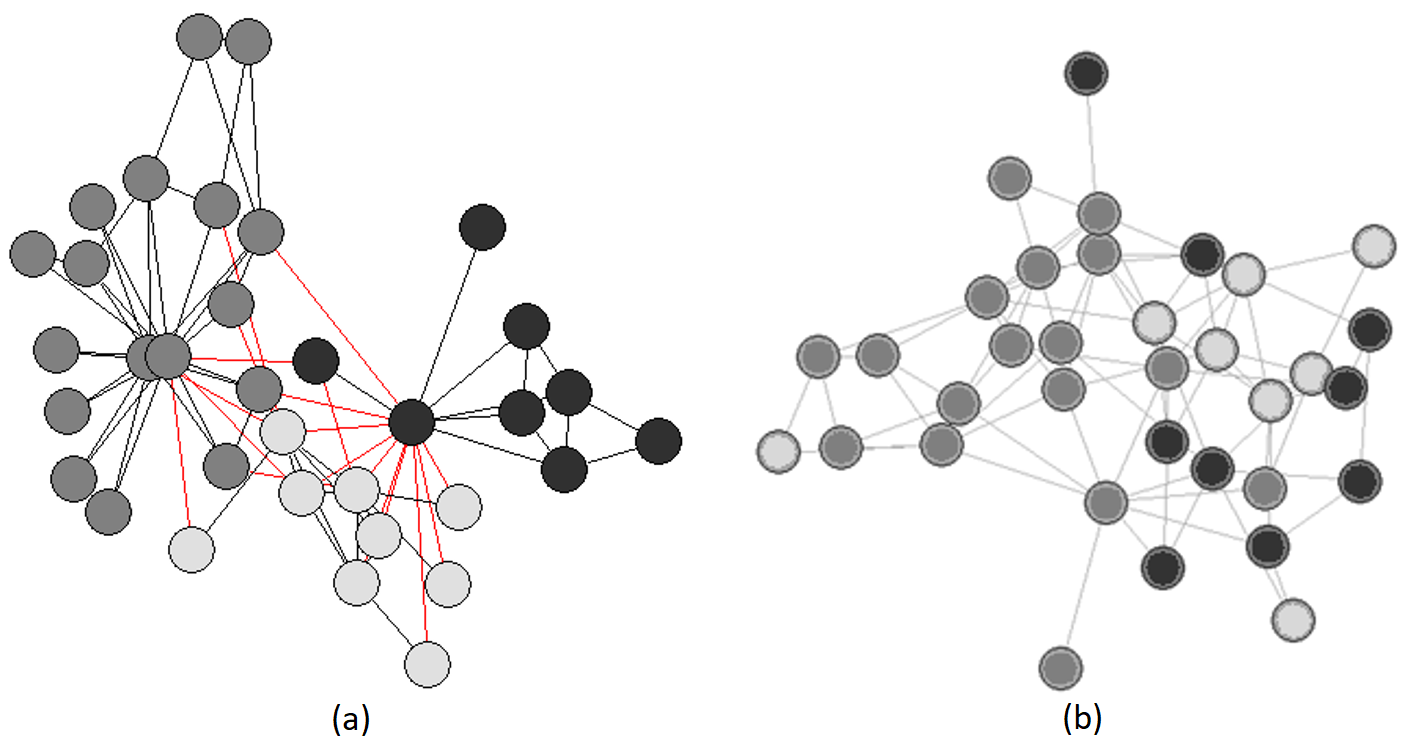}
  \caption{(a) Observed Network, (b) Generated Network}
  \label{fig:boat1}
\end{figure}

\subsection{Experiments on Caltech Network}

The following parameter values are used to generate a network that mimics Caltech network: $n = 769$, $m = 16656$, $k=8$, $C = \{0.375, 0.341, 0.254, 0.017, 0.005, 0.004, 0.003, 0.003\}$, $comp = 0.85$, $rp_n = 0.333$, $pp_n = 0.666$, $rp_e = 0.091$, $pp_e = 0.182$, $c3p_e = 0.363$, and $c4p_e = 0.363$. The experiments are repeated 10 times, and statistics for the generated and observed networks are given in Table 3.

Experimental results show that average path length, modularity, and power law exponent are closely matched with negligible standard deviations. However, clustering coefficient and diameter values are not replicated successfully in terms of difference between the generated and observed statistic. Caltech network is a larger network than the Karate network. Therefore, as desired, variances in the generated statistics are almost zero. Consequently, it can be concluded that given a sufficient number of nodes and links, our proposed growth model is able to produce stable results.

\begin{table}[h!]
\centering
\renewcommand{\arraystretch}{0.97} 
\caption{Experimental Results for Caltech Network}
\small
\setlength\tabcolsep{2.5pt}
\begin{tabular}{|l|r|r|r|r|r|} 
\cline{2-6}
\multicolumn{1}{l|}{} & \begin{tabular}[c]{@{}r@{}}Clustering\\ Coefficient \end{tabular} & \begin{tabular}[c]{@{}r@{}}Avg. Path\\ Length \end{tabular} & Modularity     & Diameter        & \begin{tabular}[c]{@{}r@{}}Power Law\\ Exponent \end{tabular}  \\ 
\hline
Run\#1                 & 0.17                                                              & 2.19                                                        & 0.32           & 4.00            & 1.52                                                           \\ 
\hline
Run\#2                 & 0.17                                                              & 2.20                                                        & 0.32           & 4.00            & 1.52                                                           \\ 
\hline
Run\#3                 & 0.17                                                              & 2.20                                                        & 0.33           & 4.00            & 1.52                                                           \\ 
\hline
Run\#4                 & 0.17                                                              & 2.19                                                        & 0.34           & 4.00            & 1.52                                                           \\ 
\hline
Run\#5                 & 0.17                                                              & 2.19                                                        & 0.32           & 4.00            & 1.52                                                           \\ 
\hline
Run\#6                 & 0.17                                                              & 2.19                                                        & 0.33           & 4.00            & 1.52                                                           \\ 
\hline
Run\#7                 & 0.17                                                              & 2.20                                                        & 0.32           & 4.00            & 1.52                                                           \\ 
\hline
Run\#8                 & 0.17                                                              & 2.20                                                        & 0.33           & 4.00            & 1.52                                                           \\ 
\hline
Run\#9                & 0.17                                                              & 2.19                                                        & 0.33           & 4.00            & 1.52                                                           \\ 
\hline
Run\#10                & 0.17                                                              & 2.20                                                        & 0.32           & 4.00            & 1.52                                                           \\ 
\hline
\textbf{Mean}         & \textbf{0.17}                                                     & \textbf{2.20}                                               & \textbf{0.33}  & \textbf{4.00}   & \textbf{1.52}                                                  \\ 
\hline
\textit{StdDev}       & \textit{0.00}                                                     & \textit{0.00}                                               & \textit{0.01}  & \textit{0.00}   & \textit{0.00}                                                  \\ 
\hline
\textbf{Observed}     & \textbf{0.29}                                                     & \textbf{2.34}                                               & \textbf{0.33}  & \textbf{6.00}   & \textbf{1.50}                                                  \\ 
\hline
\textbf{Diff.}        & \textbf{-0.12}                                                    & \textbf{-0.14}                                              & \textbf{0.00}  & \textbf{-2.00}  & \textbf{0.02}                                                  \\
\hline
\end{tabular}
\end{table}

\subsection{Other Experiments}
Another set of experiments are performed without the purpose of mimicking any real network but to explore the variety of networks which can be generated with our proposed growth model. Parameter settings for eight experiments are presented in Table 4. The first and last four experiments are the same except that $m=2000$ for former experiments whereas $m=5000$ for latter experiments, hence a denser network. Each experiment is repeated for 10 times. The mean and standard deviations of generated statistics are presented in Table 5.

\begin{table}[h!]
\centering
\renewcommand{\arraystretch}{0.97} 
\caption{Settings for Other Experiments}
\small
\setlength\tabcolsep{2.4pt}
\begin{tabular}{|l|r|r|l|r|r|r|r|r|r|r|r|} 
\cline{2-12}
\multicolumn{1}{l|}{} & \multicolumn{1}{l|}{$n$ } & \multicolumn{1}{l|}{$m$ } & $k$  & \multicolumn{1}{l|}{$C$ }         & $comp$  & $rp_n$  & $pp_n$  & \multicolumn{1}{l|}{$rp_e$ } & \multicolumn{1}{l|}{$pp_e$ } & \multicolumn{1}{l|}{$c3p_e$ } & \multicolumn{1}{l|}{$c4p_e$ }  \\ 
\hline
Exp\#1                & 500                       & 2000                      & 5    & \{.2, 0.2, 0.2, 0.2, 0.2\}        & 0.50    & 0.50    & 0.5     & 0.25                         & 0.25                         & 0.25                          & 0.25                           \\ 
\hline
Exp\#2                & 500                       & 2000                      & 5    & \{.2, 0.2, 0.2, 0.2, 0.2\}        & 0.75    & 0.33    & 0.66    & 0.10                         & 0.20                         & 0.30                          & 0.40                           \\ 
\hline
Exp\#3                & 500                       & 2000                      & 5    & \{.005, 0.055, 0.11, 0.28, 0.55\} & 0.75    & 0.33    & 0.66    & 0.10                         & 0.20                         & 0.30                          & 0.40                           \\ 
\hline
Exp\#4                & 500                       & 2000                      & 5    & \{.005, 0.055, 0.11, 0.28, 0.55\} & 0.25    & 0.17    & 0.83    & 0.12                         & 0.63                         & 0.12                          & 0.12                           \\ 
\hline
Exp\#5                & 500                       & 5000                      & 5    & \{.2, 0.2, 0.2, 0.2, 0.2\}        & 0.50    & 0.50    & 0.50    & 0.25                         & 0.25                         & 0.25                          & 0.25                           \\ 
\hline
Exp\#6                & 500                       & 5000                      & 5    & \{.2, 0.2, 0.2, 0.2, 0.2\}        & 0.75    & 0.33    & 0.66    & 0.10                         & 0.20                         & 0.30                          & 0.40                           \\ 
\hline
Exp\#7                & 500                       & 5000                      & 5    & \{.005, 0.055, 0.11, 0.28, 0.55\} & 0.75    & 0.33    & 0.66    & 0.10                         & 0.20                         & 0.30                          & 0.40                           \\ 
\hline
Exp\#8                & 500                       & 5000                      & 5    & \{.005, 0.055, 0.11, 0.28, 0.55\} & 0.25    & 0.17    & 0.83    & 0.12                         & 0.63                         & 0.12                          & 0.12                           \\
\hline
\end{tabular}
\end{table}

\begin{table}[h!]
\centering
\caption{Other Experimental Results}

\small
\setlength\tabcolsep{2.5pt}
\begin{tabular}{|l|l|r|r|r|r|r|} 
\cline{3-7}
\multicolumn{1}{l}{} &                    & \begin{tabular}[c]{@{}r@{}}Clustering\\ Coefficient \end{tabular} & \begin{tabular}[c]{@{}r@{}}Avg. Path\\ Length \end{tabular} & Modularity     & Diameter       & \begin{tabular}[c]{@{}r@{}}Power Law\\ Exponent \end{tabular}  \\ 
\hline
Exp\#1                & Mean               & 0.10                                                              & 3.31                                                        & 0.38           & 7.50           & 2.41                                                           \\ 
\cline{2-7}
                     & \textit{Std.Dev.}  & \textit{0.00}                                                     & \textit{0.02}                                               & \textit{0.01}  & \textit{0.71}  & \textit{0.03}                                                  \\ 
\hline
Exp\#2                & Mean               & 0.11                                                              & 3.33                                                        & 0.47           & 7.60           & 2.41                                                           \\ 
\cline{2-7}
                     & \textit{Std.Dev.}  & \textit{0.00}                                                     & \textit{0.03}                                               & \textit{0.01}  & \textit{0.52}  & \textit{0.04}                                                  \\ 
\hline
Exp\#3                & Mean               & 0.10                                                              & 3.29                                                        & 0.34           & 7.60           & 2.39                                                           \\ 
\cline{2-7}
                     & \textit{Std.Dev.}  & \textit{0.00}                                                     & \textit{0.02}                                               & \textit{0.02}  & \textit{0.70}  & \textit{0.03}                                                  \\ 
\hline
Exp\#4                & Mean               & 0.11                                                              & 3.39                                                        & 0.42           & 7.60           & 2.47                                                           \\ 
\cline{2-7}
                     & \textit{Std.Dev.}  & \textit{0.01}                                                     & \textit{0.02}                                               & \textit{0.01}  & \textit{0.52}  & \textit{0.04}                                                  \\ 
\hline
Exp\#5               & Mean               & 0.15                                                              & 2.51                                                        & 0.36           & 5.10           & 1.82                                                           \\ 
\cline{2-7}
                     & \textit{Std.Dev.}  & \textit{0.00}                                                     & \textit{0.00}                                               & \textit{0.01}  & \textit{0.32}  & \textit{0.01}                                                  \\ 
\hline
Exp\#6                & Mean               & 0.16                                                              & 2.52                                                        & 0.44           & 4.90           & 1.81                                                           \\ 
\cline{2-7}
                     & \textit{Std.Dev.}  & \textit{0.00}                                                     & \textit{0.01}                                               & \textit{0.00}  & \textit{0.32}  & \textit{0.01}                                                  \\ 
\hline
Exp\#7                & Mean               & 0.15                                                              & 2.51                                                        & 0.31           & 5.20           & 1.82                                                           \\ 
\cline{2-7}
                     & \textit{Std.Dev.}  & \textit{0.00}                                                     & \textit{0.01}                                               & \textit{0.01}  & \textit{0.42}  & \textit{0.01}                                                  \\ 
\hline
Exp\#8                & Mean               & 0.17                                                              & 2.55                                                        & 0.40           & 5.00           & 1.82                                                           \\ 
\cline{2-7}
                     & \textit{Std.Dev.}  & \textit{0.00}                                                     & \textit{0.01}                                               & \textit{0.02}  & \textit{0.00}  & \textit{0.01}                                                  \\
\hline
\end{tabular}
\end{table}

The experiments confirm that standard deviations are indeed very low. Diameter might be seen as an exception to this but it is mostly due to the nature of that statistic. Since it is an integer value, even if the model generates only two adjacent integer values for this statistic, standard deviation might come as large at the first sight.

As the number of links increase from 2000 to 5000, values for average path Length and diameter get smaller as expected. On the other hand, clustering coefficent increases.  There does not seem to be a meaningful change in the values of modularity.

\section{Conclusion}
In this work, we have developed a network growth model with the aim of exploring the mechanisms behind most real-world networks, and being able to mimic real-world networks through these mechanisms. The proposed growth model serves as an initial step to build a realistic growth model which can generate graphs with any set of given statistics.

When there exist a sufficient number of nodes and links, our model generates networks which do not vary between themselves in terms of the network statistics employed in this study. Such stability is highly desired. However, some network statistics are not being successfully replicated. This might be because one or combination of the two things. First, parameter values are determined based on a few manual experiments rather than estimating them from the real network. Second, the mechanisms employed in the growth model might not be sufficient to generate any social network easily, which suggest further development of the model.

Given the limitations of the current work, more effort in future should be directed towards developing more accurate mechanisms, exploring relationships between the mechanisms, and estimating parameter values via more systematic calibration experiments.

 \newpage
\section*{References}

Barabasi, A. L., \& Albert, R. (1999). Emergence of scaling in random networks. Science, 286(5439), 509-512. doi:10.1126/science.286.5439.509\vspace{4pt}

\noindent Erdos, P., \& Renyi, A. (1959). On random graphs I. Publicationes Mathematicae (Debrecen), 6, 290-297.\vspace{4pt}

\noindent Gilbert, E. N. (1959). Random graphs. The Annals of Mathematical Statistics, 30(4), 1141-1144.\vspace{4pt}

\noindent Kimura, M., Saito, K., \& Ueda, N. (2004). Modeling of growing networks with directional attachment and communities. Neural Networks, 17(7), 975-988. doi:10.1016/j.neunet.2004.01.005\vspace{4pt}

\noindent Lazega, E., \& Pattison, P. E. (1999). Multiplexity, generalized exchange and cooperation in organizations: A case study. Social Networks, 21(1), 67-90. doi:10.1016/s0378-8733(99)00002-7\vspace{4pt}

\noindent Leskovec, J., Backstrom, L., Kumar, R., \& Tomkins, A. (2008). Microscopic evolution of social networks. Proceeding of the 14th ACM SIGKDD International Conference on Knowledge Discovery and Data Mining - KDD 08. doi:10.1145/1401890.1401948\vspace{4pt}

\noindent Lim, S., Lee, S., Powers, S. S., Shankar, M., \& Imam, N. (2016). Survey of Approaches to Generate Realistic Synthetic Graphs. doi:10.2172/1339361\vspace{4pt}

\noindent Newman, M. (2005). Power laws, Pareto distributions and Zipfs law. Contemporary Physics, 46(5), 323-351. doi:10.1080/00107510500052444\vspace{4pt}

\noindent Price, D. D. (1976). A general theory of bibliometric and other cumulative advantage processes. Journal of the American Society for Information Science, 27(5), 292-306. doi:10.1002/asi.4630270505\vspace{4pt}

\noindent Simmel, G. (1908). Sociologie. Untersuchungen über die Formen der Vergesellschaftung. Duncker \& Humblot.\vspace{4pt}

\noindent Simon, H. A. (1955). On a Class of Skew Distribution Functions. Biometrika, 42(3/4), 425. doi:10.2307/2333389\vspace{4pt}

\noindent Snijders, T. A., Pattison, P. E., Robins, G. L., \& Handcock, M. S. (2006). New Specifications for Exponential Random Graph Models. Sociological Methodology, 36(1), 99-153. doi:10.1111/j.1467-9531.2006.00176.x\vspace{4pt}

\noindent Traud, A. L., Kelsic, E. D., Mucha, P. J., \& Porter, M. A. (2011). Comparing Community Structure to Characteristics in Online Collegiate Social Networks. SIAM Review, 53(3), 526-543. doi:10.1137/080734315\vspace{4pt}

\noindent Zachary, W. W. (1977). An Information Flow Model for Conflict and Fission in Small Groups. Journal of Anthropological Research, 33(4), 452-473. doi:10.1086/jar.33.4.3629752\vspace{4pt}

\end{document}